\documentclass[twocolumn,showpacs,amsmath,amssymb,superscriptaddress]{revtex4-1}
\usepackage{amsmath,amssymb,color,latexsym,natbib}
\usepackage{dcolumn}
\usepackage{bm}
\usepackage{color}
\usepackage{ulem}
\usepackage{soul}
\def\ADD#1{{\textcolor{black}{#1}}}
\newcommand{\be}{\begin{equation}}
\newcommand{\ee}{\end{equation}}
\def\ba{\begin{eqnarray}}
\def\ea{\end{eqnarray}}

\def\ie{{\it i.e.}\ }
\def\kk{{\bf k}}

\begin{document}
\preprint{1}

\title{Turbulence of Weak Gravitational Waves in the Early Universe}
\author{S\'ebastien Galtier}
\affiliation{Laboratoire de Physique des Plasmas, \'Ecole Polytechnique, Univ. Paris-Sud, Universit\'e Paris-Saclay, F-91128 Palaiseau Cedex, France}
\email{sebastien.galtier@u-psud.fr}
\author{Sergey V. Nazarenko}
\affiliation{Mathematics Institute, University of Warwick, Coventry CV4~7AL, UK}
\email{S.V.Nazarenko@warwick.ac.uk}
\date{\today}

\begin{abstract}
We study the statistical properties of an ensemble of weak gravitational waves interacting nonlinearly in a flat space-time.
We show that the resonant three-wave interactions are absent and develop a theory for four-wave interactions in the reduced case of a 
$2.5+1$ diagonal metric tensor. In this limit, where only $+$ gravitational waves are present, we derive the interaction Hamiltonian and 
consider the asymptotic regime of weak gravitational wave turbulence. Both direct and inverse cascades are found for the energy 
and the wave action respectively, and the corresponding wave spectra are derived. The inverse cascade is characterized by a finite-time 
propagation of the metric excitations -- a process similar to an explosive non-equilibrium Bose--Einstein condensation, which provides an 
efficient mechanism to ironing out small-scale inhomogeneities. The direct cascade leads to an accumulation of the radiation energy in the system.
These processes \ADD{might be} important for understanding the early Universe where a background of weak nonlinear gravitational 
waves is expected. 
\end{abstract}


\maketitle

\paragraph*{Introduction.}
The recent direct observations of gravitational waves (GW) by the LIGO-\ADD{VIRGO} collaboration \cite{LIGO}, a century after their prediction by Einstein 
\cite{einstein}, is certainly one of the most important events in astronomy which opens a new window onto the Universe, the so-called GW astronomy.
In modern Universe, shortly after being excited by a source, e.g. a merger of two black holes, GW become essentially linear and therefore 
non-interacting during their subsequent propagation. 
In the very early Universe, different mechanisms have been proposed for the generation of primordial GW, like e.g. phase 
transition \cite{witten84,krauss,kamionkowski,battye,jones,dev,jinno}, self-ordering scalar fields \cite{fenu}, cosmic strings \cite{damour} and 
cosmic defects \cite{Figueroa}. Production of GW is also expected to have taken place during the cosmological inflation era \cite{guth,rubakov,guzzetti}
and many efforts are currently made to detect indirectly their existence \cite{bicep}.
The physical origin of the exponential expansion of the 
early Universe is, however, not clearly explained and still under investigation \cite{weinberg08,marochnik}. Formally, it was incorporated into the 
general relativity equations simply through adding a positive cosmological constant.

The primordial GW were, presumably, significantly more nonlinear than the GW in the modern Universe (like the GW observed recently by LIGO-\ADD{VIRGO})
as they had much larger energy packed in a much tighter space \citep{pen}. 
\ADD{Although  not firmly validated, a scenario was suggested in which} a first-order phase transition  proceeds through the collisions of true-vacuum bubbles creating a
 potent source of GW \cite{turner,kosowsky,binetruy}.
\ADD{According to this scenario}, 
at the time of the grand-unified-theory (GUT) symmetry breaking ($t_* \sim 10^{-36}$\,sec, $T_* \sim 10^{15}$\,Gev), the ratio of the 
energy density in GW ($\rho_{GW}$) to that in radiation ($\rho_{rad}$) after the transition is about $5\%$ \cite{kosowsky}. From the expressions given 
in \cite{kosowsky} and using as a time-scale $t_*$ (and also $g_* \sim 100$), we find the following estimate for the GW amplitude: $h \sim 0.3$.
\ADD{Supposedly}, 
such waves were covering the Universe quasi-uniformly rather than being concentrated locally in space and time near an isolated burst 
event, and it is likely that their distribution was broad in frequencies and propagation angles. 
At some stage of expansion of the Universe the GW had become rather weak, but still nonlinear enough for having non-trivial mutual interactions. 
Importance of the nonlinear nature of the GW was pointed out in the past for explaining, e.g. the memory effect \cite{thorne} 
or part of the dark energy \cite{chevalier}.
The possibility to get a turbulent energy cascade of the primordial gravitons was also mentioned \cite{efroimsky,yang15} but, to date, no theory 
has been developed. A  turbulence theory seems to be particularly relevant for GW because they are nonlinear and  their dissipation is negligible. 
Recent works \cite{yang,green} explore  some ideas on similar lines: they  investigate numerically the turbulent 
nature of black holes, define a gravitational Reynolds number, and show that the system can display a nonlinear parametric instability with transfers 
reminiscent of an inverse cascade (see also Refs. \cite{carrasco,adams}).

The nonlinear properties of the GW, especially the primordial GW \ADD{mentioned above},
call for using the wave turbulence approach considering statistical behaviour  of random  weakly nonlinear  waves  
\citep{ZLF,naza11}. The energy transfer between such waves occurs mostly within resonant sets of waves and the resulting energy distribution, 
far from a thermodynamic equilibrium, is often characterized by exact power law solutions similar to the Kolmogorov spectrum of 
hydrodynamic turbulence -- the so-called Kolmogorov--Zakharov (KZ) spectra \citep{ZLF,naza11}. 
The wave turbulence approach has  been successfully applied to many diverse physical systems like, e.g. capillary and gravity waves 
\citep{deike,denis,denis1,Falcon,aubourg}, superfluid helium and processes of Bose-Einstein condensation \citep{lvov03}, nonlinear optics 
\citep{Dyachenko}, rotating fluids \citep{Galtier2003}, geophysics \citep{Galtier2014}, elastic waves \cite{chibbaro} or astrophysical plasmas 
\citep{space} (see \citep{naza11} for a more detailed list of references). 

In this Letter, we develop a theory of weak GW turbulence at the level of four-wave interactions in a reduced setup of a $2.5+1$ diagonal metric tensor. 
\ADD{The physical properties of such a system are first rigorously derived. Then, in the last section, we present a non rigorous discussion of a potential 
connection to the physics of the very early Universe.}

\paragraph*{Absence of resonant three-wave interactions.}
We shall consider Einstein's general relativity equations (free of the cosmological constant) for an empty space, $R_{\mu\nu} =0$, where $R_{\mu\nu}$ is 
the Ricci curvature tensor. We will be interested in weak space-time ripples on the background of a flat space. 
Respectively, the  metric tensor will be assumed to have the form
$g_{\mu \nu} = \eta_{\mu \nu} +  h_{\mu \nu}$, where $h_{\mu \nu} \ll 1$ and $\eta_{\mu \nu}$ is the Poincar\'e-Minkowski flat space-time 
 metric. In the linear approximation with the gauge conditions, Einstein's vacuum equations give rise to two GW modes: the $+$ and 
$\times$--polarized ones \citep{maggiore}. Next order in small wave amplitudes leads to terms with quadratic nonlinearities which are often associated 
with triadic resonant interactions. To describe such triadic interactions of GW we need to consider the quadratic part of the Ricci tensor, 
$R_{\mu\nu} =  R^{(1)}_{\mu\nu} +  R^{(2)}_{\mu\nu}$ with $R^{(1)}_{\mu\nu} = - \Box h_{\mu\nu}$ and 
\citep{weinberg}
\ba \label{eqs:hint5}
R^{(2)}_{\mu \nu}  =  + \frac 14
\left[2 \frac {\partial h^\alpha_{\sigma}}{ \partial x^\alpha}
- \frac {\partial h^\alpha_{\alpha}}{ \partial x^\sigma}
\right]
\left[\frac {\partial h^\sigma_{\mu}}{ \partial x^\nu}
+ \frac {\partial h^\sigma_{\nu}}{ \partial x^\mu}
-  \frac {\partial h_{\mu \nu}}{ \partial x_\sigma}
\right] \quad \quad \\
-\frac 12
h^{\lambda \alpha}  \left[ \frac {\partial^2 h_{\lambda \alpha}}{\partial x^\nu \partial x^\mu}
- \frac {\partial^2 h_{\mu \alpha}}{\partial x^\nu \partial x^\lambda}
-\frac {\partial^2 h_{\lambda \nu}}{\partial x^\alpha \partial x^\mu}
+\frac {\partial^2 h_{\mu \nu}}{\partial x^\alpha \partial x^\lambda}
\right]
 \nonumber \\
- \frac 14
\left[ \frac {\partial h_{\sigma\nu}}{ \partial x^\lambda}
+ \frac {\partial h_{\sigma\lambda}}{ \partial x^\nu}
- \frac {\partial h_{\lambda\nu}}{ \partial x^\sigma}
\right]
\left[\frac {\partial h^\sigma_{\mu}}{ \partial x_\lambda}
+ \frac {\partial h^{\sigma \lambda}}{ \partial x^\mu}
-  \frac {\partial h^\lambda_{\mu }}{ \partial x_\sigma}
\right] . \nonumber
\ea
Weak turbulence theory predicts that  resonant n-wave interactions play the dominant role for the nonlinear evolution. For the three-wave interactions we have  the conditions 
$\kk = \kk_1 + \kk_2$ and $ \omega_{\bf k} =  \omega_{\kk_1} +  \omega_{\kk_2}$, with the dispersion relation $\omega_{\bf k} = c \vert \kk \vert = c k$, 
where $c$ is the speed of light ($c=1$ thereafter) and $\kk$ is the wave vector.
These resonant conditions are formally identical to the respective conditions for the acoustic wave turbulence problem for which it is well-known that all the resonant triads consist of collinear $\kk$'s \cite{ZLF}.
Therefore, in the physical space  the three-wave resonant interactions
split the  3D dynamics into individual 1D systems independent for all particular directions. Let us choose one of such directions, and let  our $z$-axis to be parallel to the chosen 
direction. We shall use the transverse-traceless gauge, \ie $h^\mu_\mu=0$, $\partial_\mu h_{\mu \nu}=0$ and $h^{0\nu}=0$ \citep{maggiore}. 
Then, the normal mode structure is $h^+_{1 1} = -h^+_{2 2}=a$ corresponding to the $+$ GW and $h^\times_{1 2} = h^\times_{2 1} = b$
corresponding to the $\times$ waves (all the other tensor components are zero). 
Evolution equations for $a$ and $b$ follow from taking the respective projections in  equation 
$\Box h_{\mu \nu} = 2  R^{(2)}_{\mu\nu}$, which gives $\Box a =  R^{(2)}_{11} -  R^{(2)}_{22}$ 
and $\Box b =  R^{(2)}_{12} +  R^{(2)}_{21}$. 
Substitute here the respective components of $R_{\mu \nu}^{(2)}$ from expression (\ref{eqs:hint5}) in which only derivatives with respect 
to $t$ and $z$ are left; this gives after some calculations (we define $\dot{a} = \partial_t a$ etc.)
$R^{(2)}_{11} = R^{(2)}_{22} = \frac{1}{2} [ \dot{a}^2 + \dot{b}^2 - \left(\partial_z a\right)^2 - \left(\partial_z b\right)^2]$, 
and $R^{(2)}_{12} = R^{(2)}_{21} =0$, so that $\Box a = 0$ and $\Box b = 0$.
Therefore three-wave interactions of weak GW are absent, and the dominant resonant interactions in weak GW turbulence is four-wave or higher.

\paragraph*{Theory for four-wave interactions.}
To calculate the four-wave interactions, one has to expand the Einstein's equations up to the third-order nonlinearity and perform a canonical 
transformation to eliminate the quadratic nonlinearities. In the general case, this seems to be a  laborious task, dealing with which we postpone to 
future. In this Letter, we will simplify our treatment of interacting GW by considering a $2.5+1$ diagonal reduction recently studied in the framework 
of strong GW \citep{HZ14}. 
This is probably the simplest metric that contains  nonlinear properties sufficient for deriving a nontrivial wave turbulence theory of random 
weakly nonlinear GW engaged in four-wave interactions.
In the past, diagonal metrics 
were used for describing a wide range of phenomena like e.g. the Schwarzschild black hole \citep{schw} or the 
Friedmann-Robertson-Walker model of cosmology \citep{weinberg}. Note that this reduced form (with fields depending on two space variables only 
but have non-zero components for the third spatial direction) is different from the $2+1$ case which does not support GW \citep{witten}.  

Let us consider the vacuum space-time evolution described by the diagonal metric tensor  \citep{HZ14}
\begin{equation}\label{I1}
g_{\mu \nu} = 
\left( \begin{array}{cccc}
-(H_0)^2 & 0 & 0 &0 \\
 0 & (H_1)^2 & 0 &0 \\
 0 & 0 & (H_2)^2  &0 \\
 0 & 0 &0 &(H_3)^2 \end{array} \right) ,
\end{equation}
where Lam\'{e} coefficients $H_0, H_1, H_2$ and $ H_3$ are functions of $x_0 =t, x_1=x$ and $x_2=y$, and independent of $x_3=z$.  
Corresponding $2.5+1$ vacuum Einstein's equations were recently proven to be compatible in a sense that the dynamics preserves the assumed form of the metric tensor \citep{HZ14}. This  provides us with a significantly simplified setup for the description of GW. 
The simplification comes at a cost: only $+$ polarized and not $\times$ polarized GW are included in the description. 
The cross-polarized waves are absent initially and are not excited during the evolution. Also, in this framework we are restricted to a 2D dependence of the physical 
space variables. However, the $2.5+1$ dynamical vacuum system appears to be a good starting point for studying the properties of interacting  GW 
and developing a wave turbulence theory. Following \cite{HZ14}, we further define, 
\begin{eqnarray}
H_0 = e^{-\lambda} \gamma, \,\,\, H_1 = e^{-\lambda} \beta, \,\,\, H_2 = e^{-\lambda} \alpha, \,\,\, H_3 = e^{\lambda}. \quad
\end{eqnarray}
In terms of  fields $\alpha,\beta,\gamma$ and $\lambda$, Einstein's equations $R_{01}=R_{02}=R_{12}=0$ and $R_{\mu \mu}=0$ 
become respectively
\ba
\label{eq:alpha}
\partial_x \dot \alpha &=& -2\alpha \dot \lambda (\partial_x \lambda) + \frac{\dot \beta (\partial_x \alpha)}{\beta} + \frac{\dot \alpha (\partial_x \gamma)}{\gamma} ,  \nonumber \\
\label{eq:beta}
\partial_y \dot \beta &=& -2 \beta \dot \lambda (\partial_y \lambda) + \frac{\dot \alpha (\partial_y \beta)}{\alpha} + \frac{\dot \beta (\partial_y \gamma)}{\gamma} ,   \nonumber \\
\label{eq:gamma}
\partial_x \partial_y \gamma &=& -2\gamma (\partial_x \lambda) (\partial_y \lambda) + \frac{(\partial_x \alpha) (\partial_y \gamma)}{\alpha} \nonumber
+ \frac{(\partial_x \gamma) (\partial_y \beta)}{\beta} \nonumber
\ea
and
\be
\partial_t { \left( \frac{\alpha\beta}{\gamma} \dot{\lambda} \right) } 
- \partial_x \left(\frac{\alpha\gamma}{\beta} \partial_x \lambda\right)-\partial_y \left(\frac{\beta \gamma}{\alpha} \partial_y \lambda \right)=0 . \nonumber
\ee
In the linear approximation, we have $\alpha=\beta=\gamma =1$ and 
$\ddot \lambda-\partial_{xx} \lambda - \partial_{yy} \lambda=0$.
This equation has a wave solution
$\lambda = c_1 \exp({-i \omega_{\bf k} t +i {\bf k} \cdot {\bf x}}) + c_2 \exp({i \omega_{\bf k} t +i {\bf k} \cdot {\bf x}})$, 
where ${\bf k} =(p,q)$ is a 2D wave vector whereas $c_1$ and $c_2$ are arbitrary constants. 

Let us introduce the perturbed variables $\tilde \alpha = \alpha -1$, $\tilde \beta = \beta -1$ and $\tilde \gamma = \gamma -1$. We can see from the 
Einstein's equations that the leading order of each of these perturbations is quadratic in the wave amplitude $\lambda$ which is of order $\epsilon$. 
Thus, in the leading order we obtain
\begin{eqnarray}
\label{eq:alphal}
\partial_x \dot{\tilde\alpha} &=& -2 \dot \lambda (\partial_x \lambda)  ,  \\
\partial_y \dot{\tilde\beta} =  -2  \dot \lambda (\partial_y \lambda) , &&
\partial_x \partial_y \tilde\gamma = -2 (\partial_x \lambda) (\partial_y \lambda)  \nonumber
\end{eqnarray}
and
\ba
\label{eq:LambdaNL}
\partial_t{ \left[( 1+\tilde\alpha+\tilde\beta- \tilde\gamma) \dot \lambda \right] } \, 
= \quad \quad\quad\quad\quad\\
 \partial_x  \left[ ( 1+\tilde\alpha-\tilde\beta +\tilde\gamma) \partial_{x} \lambda
\right] +
\partial_y  \left[ ( 1-\tilde\alpha+\tilde\beta + \tilde\gamma) \partial_{y} \lambda
\right] . \nonumber
\ea
One can obtain our dynamical equations from a variational principle for the so-called Einstein-Hilbert action defined by the Lagrangian density \cite{HZ14}
\begin{eqnarray} 
\mathcal{L} &=& \frac{1}{2} \left[\frac{\alpha \beta}{\gamma} {\dot \lambda}^2 - \frac{\alpha \gamma}{\beta} (\partial_x \lambda)^2 
- \frac{\beta \gamma}{\alpha} (\partial_y \lambda)^2  \right. \nonumber \\ 
&& \left. \hspace{0.4cm} - \frac{\dot\alpha \dot \beta}{\gamma} + \frac{(\partial_x \alpha) (\partial_x \gamma)}{\beta} + 
\frac{(\partial_y \beta)(\partial_y \gamma)}{\alpha}\right] \nonumber \\
&\approx& \mathcal{L}_ \text{free}+ \mathcal{L}_\text{int} , \, \, \, 
\label{eq:LagrangianDiagonal}
\end{eqnarray}
where $\mathcal{L}_\text{free} = \frac{1}{2} \left[ {\dot \lambda}^2 -  (\nabla \lambda)^2 \right]$
and
\begin{eqnarray} \label{eq:L2}
&&\mathcal{L}_\text{int} =
 \frac 12 \left[( \tilde\alpha +\tilde\beta - \tilde\gamma) {\dot \lambda}^2 + (- \tilde\alpha+\tilde\beta- \tilde\gamma) (\partial_x \lambda)^2  \right.  \nonumber \\ 
&& \left. +( \tilde\alpha-\tilde\beta - \tilde\gamma) (\partial_y \lambda)^2 
-{\dot {\tilde \alpha} \dot {\tilde \beta}} + {(\partial_x \tilde \alpha) (\partial_x \tilde \gamma)} + {(\partial_y \tilde \beta)(\partial_y \tilde \gamma)} \nonumber
\right],
\end{eqnarray}
representing the linear (free wave) dynamics and the (leading order of) the wave interaction respectively.
Let us deal with fields which are periodic with period $L$ in both $x$ and $y$
(limit $L\to +\infty$ to be taken later), and introduce Fourier coefficients, 
$\lambda_{\bf k}(t) =L^{-2} \int_\text{square} \lambda({\bf x},t) \, \exp(-i {\bf k} \cdot {\bf x}) dxdy$, etc.
Then
\be
\int \mathcal{L}_\text{free} \, d{\bf x} =
\frac 12  \sum_{\bf k} \left( |{\dot \lambda}_{\bf k}|^2 +k^2 |\lambda_{\bf k}|^2 \right) .
\ee
We introduce the normal variables as
\begin{eqnarray} \label{eqs:ak}
\lambda_{\bf k}  = 
\frac{a_{\bf k} + a^*_{-\bf k}}{\sqrt{2k}}, \quad
{\dot \lambda}_{\bf k}  = 
\frac{\sqrt{k}(a_{\bf k} - a^*_{-\bf k})}{i\sqrt{2}} ,
\end{eqnarray}
so that 
\ba
\int \! \mathcal{L}_\text{free} \, d{\bf x} \, dt &=&
\frac i 2  \int  \!  dt \sum_{\bf k} 
\left(a_{\bf k}^* \dot a_{\bf k} - a_{\bf k} \dot a_{\bf k}^* 
 \right) - \int  \!  H_\text{free}  \, dt , \quad \quad \nonumber \\
\int \mathcal{L}_\text{int} \, d{\bf x} \, dt &=&
- \int H_\text{int} \, dt , 
\nonumber
\ea
where 
\be
H_\text{free} = \sum_{\bf k}  k  |a_{\bf k}|^2
\ee
and
\ba \label{eq:hamiltint}
&&H_\text{int} = \frac 12 \sum_{1,2,3} \delta_{123} \left\{ (-\tilde\alpha_1 -\tilde\beta_1 + \tilde\gamma_1){\dot \lambda}_2 {\dot \lambda}_3  
-   \right.  \nonumber \\
&&\left. \left[ ( \tilde\alpha_1 -\tilde\beta_1 + \tilde\gamma_1) p_2 p_3 + ( -\tilde\alpha_1 +\tilde\beta_1 + \tilde\gamma_1) q_2 q_3 \right] \lambda_2 \lambda_3 \right\} 
\nonumber \\
&&
+\frac 12 \sum_{\bf k} 
\left[ {\dot {\tilde \alpha}}_{\bf k}{\dot {\tilde \beta}}_{\bf k}^* - (p^2 { \tilde \alpha}_{\bf k} + q^2 { \tilde \beta}_{\bf k}) {\tilde \gamma}_{\bf k}^*
\right],
\ea
are the free and  interaction Hamiltonians respectively. Here, we use shorthand notations
$ \sum_{1,2,3} =  \sum_{{\bf k}_1,{\bf k}_2,{\bf k}_3}, \delta_{123} = \delta_{{\bf k}_1+{\bf k}_2+{\bf k}_3}$ (Kronecker delta), $ \lambda_1 = \lambda_{{\bf k}_1},$  etc. 

Now we are ready to pass to the Hamiltonian description.
Taking variation of the action with respect to $a_{\bf k}^*$, we have the required Hamiltonian equation:
\be
i\dot a_{\bf k} = \frac{\partial H } {\partial{a_{\bf k}^*}}
\quad \hbox{where} \quad  H=H_\text{free} + H_\text{int} \,. \nonumber
\ee
In the linear approximation, when $H_\text{int}$ is neglected, we have the free GW solution, $a_{\bf k} \sim \exp(-ik t)$.
To find  $H_\text{int}$, in addition to expressing ${ \lambda}_{\bf k}$ and ${\dot \lambda}_{\bf k}$ in terms of ${a}_{\bf k} $ and ${a}_{\bf k}^* $, 
we have to express there ${\tilde \alpha}_{\bf k}, {\tilde \beta}_{\bf k}, {\dot {\tilde \alpha}}_{\bf k}, {\dot {\tilde \beta}}_{\bf k} $ and ${\tilde \gamma}_{\bf k} $
in terms of the same normal variables. This can be easily done  in the Fourier space (see the supplementary material).
After the introduction of these expressions and relations (\ref{eqs:ak}) into Eq.~(\ref{eq:hamiltint}),
we  obtain $H_\text{int}$ in terms of variables $a_{\bf k}$ and $a_{\bf k}^*$.
All terms in   $H_\text{int}$ are quartic in $a_{\bf k}$ and $a_{\bf k}^*$, which indicates that the leading order interaction 
process is four-wave. The terms with products of four $a_{\bf k}$'s or four $a_{\bf k}^*$'s can be dropped as they correspond to an empty 
$4 \to 0$ process. The remaining terms can be grouped into two parts:
$H_\text{int} = H_{3 \to 1} +H_{2 \to 2}$. Part $ H_{3 \to 1}$ contains products of three $a_{\bf k}$ and one $a_{\bf k}^*$ and 
vice versa -- these represent a $3 \to 1$ process. Part $H_{2 \to 2}$ contains products of two $a_{\bf k}$ and two $a_{\bf k}^*$ -- these represent a $2 \to 2$ 
process.  
Let us first consider the $3 \to 1$ process. The $3 \to 1$ resonance conditions are satisfied only by
wave quartets which are collinear (for the same reason as in the $2 \to 1$ process; see also \citep{ZLF}). Thus in this case 
we can consider contributions to the Hamiltonian from the resonant manifold only, where the quartets are collinear, which drastically
simplifies the calculation (e.g. ${p_5}/{p_1} - {q_5}/{q_1}=0$  etc.). Then, by a straightforward but lengthy calculation 
(see the supplementary material) we  find that all the $3 \to 1$ terms cancel (on the $3 \to 1$ resonant manifold), \ie $H_{3 \to 1} =0$, whereas for
 the $2 \to 2$ process we obtain  the following expression,
\be \label{kine0}
H_{2\to 2} = \sum_{1,2,3,4}  T^{12}_{34} \delta^{12}_{34} a_1 a_2 a^*_{3} a^*_{4} , 
\ee
with $T^{12}_{34} = \frac14(W^{12}_{34} + W^{21}_{34} +W^{12}_{43}+W^{21}_{43})$, $W^{12}_{34} = Q^{12}_{34} + Q^{34}_{12}$
and 
{\begin{widetext}
\ba \label{eqs:hint4}
&& \quad \quad \quad \quad Q^{12}_{34} = \frac 1{{4} {\sqrt{k_1 k_2 k_3k_4 }}} 
\left\{  
2 \left( \frac{p_{4}}{p_1-p_3} - \frac{q_{4}}{q_1-q_3} \right) \frac{k_{2}(p_1p_3-q_1q_3)}{k_1-k_3}
- 2 \left( \frac{{p_{4}}}{p_1-p_3} + \frac{{q_{4}}}{q_1-q_3} \right) \frac{k_1 k_{2} k_3} {{k_1}-k_3} \right.  \label{Q} \\
&&\left. 
+\left( \frac{p_2}{p_1+p_2} - \frac{q_2}{q_1+q_2} \right) \frac{k_1(p_3p_4-q_3q_4)} {k_1+k_2} 
- \left( \frac{p_2}{p_1+p_2} + \frac{q_2}{q_1+q_2} \right) \frac{k_1 k_3 k_4} {k_1+k_2} 
+ { \frac{2 k_1 k_3 p_2 q_4}{(p_1+p_2)(q_1+q_2)}} +  \frac{2 k_1p_3 (q_2k_4+k_2q_4)}{(p_1-p_3)(q_1-q_3)}
\right\} . \nonumber
\ea
\end{widetext}}

\noindent
Given  the standard form  of the interaction Hamiltonian (\ref{kine0}), derivation of the kinetic equation (KE) of weak wave turbulence is 
straightforward and can be found, e.g., in book \cite{naza11}, Chapter 6. The result is
\begin{widetext}
\begin{eqnarray}
 \dot n_\kk 
\label{ke}
= 4 \pi   \int |T_{{\bf k}_1{\bf k}_2}^{{\bf k}{\bf k}_3}|^2 \, 
 n_{{\bf k}_1} n_{{\bf k}_2} n_{{\bf k}_3} n_{{\bf k}} \left[
\frac{1}{n_{{\bf k}}}+\frac{1}{n_{{\bf k}_3}}-\frac{1}{n_{{\bf k}_1}} -\frac{1}{n_{{\bf k}_2}}
\right] 
\delta({ \bf k}+{ \bf k}_3-{ \bf k}_1-{ \bf k}_2) \, 
\delta(\omega_{ \bf k}+\omega_{{ \bf k}_3}-\omega_{{ \bf k}_1}-\omega_{{ \bf k}_2}) \, d { \bf k}_1 d { \bf k}_2  d { \bf k}_3 ,
\label{ke-2to2} 
\nonumber
\end{eqnarray}
\end{widetext}
where the wave action spectrum is defined as 
\be
n_{\bf k} = \lim_{L\to\infty}\frac {L^2}{4 \pi^2} \langle |a_{\bf k}|^2\rangle ,
\ee
and where $\langle \, \rangle$ denotes the ensemble average.
It is worth reminding that the KE is valid under assumptions of small nonlinearity (in our case $h \ll 1$), random phases of waves and taking 
the infinite box limit while keeping the mean wave energy density constant. 
Assuming the mirror symmetry of the spectrum $n_{\bf k} = n_{-{\bf k}}$, we have in terms of the original variables
$n_{\bf k} = k \lim_{L\to\infty}\frac {L^2}{4 \pi^2} \langle |\lambda_{\bf k}|^2\rangle \sim h^2 \ell $, where $h$ is the typical size of the metric 
ripples and $\ell$ is the typical length-scale. 
The KE has the following isotropic constant-flux stationary KZ solutions (see e.g. Eqs. (9.36) and (9.37) in book \cite{naza11} reproduced 
via a dimensional derivation in the supplementary material)
\be
n_{\bf k} \sim k^{-2} \quad \text{and} \quad n_{\bf k} \sim k^{-5/3},
\label{KZS}
\ee
corresponding respectively to the direct cascade of the vacuum ripple energy from small to large $k$'s, and  to the inverse cascade of the wave action 
("number of gravitons") from large to small $k$'s. 
Extension to the 3D isotropic geometry (also given in the supplementary material) gives $n_{\bf k} \sim k^{-3} \quad \text{and} \quad n_{\bf k} \sim k^{-8/3}$.
There is also a solution corresponding to thermodynamic equilibrium (in any geometry), the Rayleigh-Jeans spectrum, $n_{\bf k} =T/(k+\mu),$ ($T,\mu=$const).

Interestingly the 1D spectra (\ref{KZS}) can be recovered with a simple phenomenology. 
For four-wave interactions, the typical time-scale of the cascade is $\tau_{cas} \sim \tau_{GW} / \epsilon^4$, where the small parameter,
$\epsilon \sim \tau_{GW} / \tau_{NL} \ll 1$, measures the time-scale separation between the wave period $\tau_{GW} \sim 1/\omega$ and the nonlinear
time $\tau_{NL} \sim \ell / (hc)$, which follows from the perturbed  Ricci tensor. 
The KE conserves the total energy ${\cal E} = \int \omega_{ \bf k}  n_{ \bf k} d { \bf k}$ and the total wave action 
${\cal N} = \int  n_{ \bf k} d { \bf k} $ (per unit area). 
Let us consider the energy ${\cal E}_\ell$ within the scales greater than $\ell$, namely ${\cal E}_\ell = \int_{k'<k} E^{(1D)}_{k'} \, dk'$, where 
we have introduced the 1D energy spectrum $E^{(1D)}_k$.
We have ${\cal E}_\ell \sim \frac{c^4}{32 \pi G} \frac{h^2}{\ell^2}$ \cite{maggiore}, so that 
\be
\varepsilon \sim \frac{{\cal E}_\ell}{\left( \frac{\tau_{NL}}{\tau_{GW}} \right)^3 \tau_{NL}}
\sim \frac{{\cal E}_\ell}{\left( \frac{\ell}{h} \right)^4 \omega^3}
\sim \frac{{\cal E}_\ell^3}{k^3} \sim {E_k^{(1D)}}^3 \, , 
\ee
where $\varepsilon$ is the constant energy flux. This gives  the spectrum $E^{(1D)}_k \sim \varepsilon^{1/3} k^0$. 
With the wave action flux $\zeta$, we obtain in the same manner $N^{(1D)}_k \sim \zeta^{1/3} k^{-2/3}$.

\paragraph*{Discussion.}
Which of the two KZ spectra is more relevant depends on the position of the forcing scale of the space-time ripples with respect to the available range of scales 
for the GW. In turn, this may depend on the stage of the Universe evolution and on the respective physical processes generating the GW at that particular epoch.
\ADD{As discussed in the introduction, a first-order phase transition could possibly provide an efficient mechanism to generate GW with fairly high energy 
\cite{turner,kosowsky,binetruy} 
at a critical time in the very early Universe. The validity of this scenario for our Universe is, however, still not clear and needs further investigations that could be 
done in the future. Having in mind these limitations, we can still discuss the potential consequences of presence of a weak GW turbulence in the very early Universe.}
Before the  inflation era, when all  observable today parts of the Universe were within the horizon, the inverse cascade may have been an effective 
mechanism for ``ironing out" the small-scale inhomogeneities, i.e. correlating the motions of the causally connected parts of the Universe.
Indeed, this spectrum has a finite capacity: the integral defining the total wave action, $\int n_{\bf k} d{\bf k}$, converges at ${\bf k} \to 0$. 
Such systems admit self-similar solutions of the second kind in which the spectral front traverses (explosively) the infinite range of scales in a finite time 
\cite{space,colm}. In our case this means that the largest length-scale of the system will be excited by the GW spectrum in a finite time 
$t_*$ -- a kind of non-equilibrium Bose-Einstein condensation which was previously studied in the Gross-Pitaevskii model \citep{lacaze,naza11}. 
Based on the KE and putting back the original physical constants, we can estimate $t_* \sim \ell / (ch^4) $.
Of course, the mode $k=0$ (the infinite scale) is never reached.
Indeed, the weak turbulence KE  fails when the time ratio $\chi \equiv \tau_{GW} / \tau_{NL} \sim h$ becomes of order unity. By using the wave action spectrum 
we find $\chi \sim \ell^{1/3}$ showing the existence of a maximal scale $\ell_{max}$ beyond which turbulence becomes strong. 
The dilution of the GW energy due to the expansion of the Universe provides another constraint for $\ell_{max}$. 
First of all we have a restriction at the linear level: the GW cannot be longer than the Hubble horizon distance $d=c/H$. On the nonlinear level, the expansion acts as
an effective large-scale dissipation, which arrests  the inverse cascade at $k_{min} \sim 1 / (h^4 d)$  (see the supplemental material). The inverse cascade range forms 
if this wave number is greater than the forcing scale.

The direct cascade of GW is of the infinite capacity type, i.e. it corresponds to a growth of the GW physical-space energy density, until this process is saturated 
at the value ${\cal E} \sim {d h_0^6 c^4 } / (\ell_0^3 G)$ due to the expansion of the Universe (see the supplemental material).
Here, $h_0$ and $\ell_0$ refer to the values of the metric disturbance and their typical length at the forcing scale.
Such a  GW energy \ADD{may} play an important role for the overall expansion rate at the late stages of the inflation and the transition to the radiation 
dominated Universe. One could describe this effect in future by combining the coarse-graining method developed in \cite{chevalier}
with the wave turbulence approach developed in our work.

Production of a GW background is a fundamental prediction of any cosmological inflationary model \cite{guzzetti}. The features of such a fossil 
signal encode unique information about the physics of the early Universe that might be detected in the future \cite{caprini,janssen}. Here, we predict 
the form of the GW spectra emerging from random nonlinear interactions. 
Our theory is, \ADD{however}, not strictly limited to the very early Universe. For example, turbulent black holes are potentially another application: our scenario
based on 4-wave interactions can be the explanation of the inverse cascade observed recently in \cite{yang,green,carrasco,adams}.
Finally, we point out that in our case the dual cascade behavior is caused by the fact that the leading-order interaction is four-wave. We remind that absence of three-wave 
interactions was proven for the most general 3+1 metric and, therefore, we expect existence of the inverse cascade solution in such a general case too.


\begin{thebibliography}{}
\bibitem{LIGO}
B.P. Abbott et al., Phys. Rev. Lett. {\bf 116}, 061102 (2016); B.P. Abbott et al., Phys. Rev. Lett. {\bf 116}, 241103 (2016); \ADD{B.P. Abbott et al., Phys. Rev. Lett. {\bf 119}, 141101 (2017).} 
\bibitem{einstein}
A. Einstein, Sitzungsber K. Preuss Akad. Wiss. {\bf 1}, 688 (1916).
\bibitem{witten84}
E. Witten, Phys. Rev. D {\bf 30}, 272 (1984). 
\bibitem{krauss}
L.M. Krauss, Phys. Lett. B {\bf 284}, 229 (1992). 
\bibitem{kamionkowski}
M. Kamionkowski, A. Kosowsky, and M.S. Turner, Phys. Rev. D {\bf 49}, 2837 (1994). 
\bibitem{battye}
R.A. Battye, and E.P.S. Shellard, Class. Quantum Grav. {\bf 13}, A239 (1996). 
\bibitem{jones}
K. Jones-Smith, L.M. Krauss, and H. Mathur, Phys. Rev. Lett. {\bf 100}, 131302 (2008). 
\bibitem{dev}
P.S. Bhupal Dev, and A. Mazumdar, Phys. Rev. D {\bf 93}, 104001 (2016). 
\bibitem{jinno}
R. Jinno, and M. Takimoto, Phys. Rev. D {\bf 95}, 024009 (2017). 
\bibitem{fenu}
E. Fenu, D.G. Figueroa, R. Durrer, and J. Garcia-Bellido, J. Cosm. Astrop. Phys. {\bf 10}, 005 (2009). 
\bibitem{damour}
T. Damour, and A. Vilenkin, Phys. Rev. Lett. {\bf 85}, 3761 (2000);
T. Damour, and A. Vilenkin, Phys. Rev. D {\bf 71}, 063510 (2005).
\bibitem{Figueroa}
D.G. Figueroa, M. Hindmarsh, and J. Urrestilla, Phys. Rev. Lett. {\bf 110} 101302 (2013). 
\bibitem{guth}
A.H. Guth, Phys. Rev. D {\bf 23}, 347 (1981).
\bibitem{rubakov}
V.A. Rubakov, M.V. Sazhin, and A.V. Veryaskin, Phys. Lett. {\bf 115B} 189 (1982). 
\bibitem{guzzetti}
M.C. Guzzetti, N. Bartolo, M. Liguori, and S. Matarrese, Nuovo Cimento Rivista Serie {\bf 39}, 399 (2016). 
\bibitem{bicep}
P.A.R. Ade et al., Phys. Rev. Lett. {\bf 114} 101301 (2015). 
\bibitem{weinberg08}
S. Weinberg, {\em Cosmology}, (Oxford Univ. Press, New York, 2008). 
\bibitem{marochnik}
L. Marochnik, Gravit. Cosm. {\bf 22} 10 (2016). 
\bibitem{pen}
U.-L. Pen, and N. Turok, Phys. Rev. Lett. {\bf 117}, 131301 (2016). 
\bibitem{turner}
M.S. Turner, and F. Wilczek, Phys. Rev. Lett. {\bf 65} 3080 (1990). 
\bibitem{kosowsky}
A. Kosowsky, M.S. Turner, and R. Watkins, Phys. Rev. Lett. {\bf 69}, 2026 (1992). 
\bibitem{binetruy}
P. Bin\'etruy, A. Boh\'e, C. Caprini, and J.-F. Dufaux, J. Cosm. Astropart. Phys. {\bf 6}, 27 (2012). 
\bibitem{thorne}
K.S. Thorne, Phys. Rev. D {\bf 45} 520 (1992). 
\bibitem{chevalier}
C. Chevalier, F. Debbasch, and Y. Ollivier, Nonlinear Analysis {\bf 71}, e199 (2009).
\bibitem{efroimsky}
M. Efroimsky, Phys. Rev. D {\bf 49} 6512 (1994). 
\bibitem{yang15}
H. Yang, F. Zhang, S.R. Green, and L. Lehner, Phys. Rev. D {\bf 91} 084007 (2015).
\bibitem{yang}
H. Yang, A. Zimmerman, and L. Lehner, Phys. Rev. Lett. {\bf 114}, 081101 (2015). 
\bibitem{green}
S.R. Green, F. Carrasco, and L. Lehner, Phys. Rev. X {\bf 4}, 011001 (2014). 
\bibitem{carrasco}
F. Carrasco, L. Lehner, R.C. Myers, O. Reula, and A. Singh, Phys. Rev. D {\bf 86}, 126006 (2012). 
\bibitem{adams}
A. Adams, P.M. Chesler, and H. Liu, Phys. Rev. Lett. {\bf 112}, 151602 (2014). 
\bibitem{naza11}
S.V. Nazarenko, {\em Wave Turbulence}, (Lecture Notes in Physics, Berlin Springer Verlag, 2011).
\bibitem{ZLF}
V.E. Zakharov, V. L'vov, and G.E. Falkovich, {\em Kolmogorov Spectra of Turbulence}, (Springer, Berlin, 1992).
\bibitem{deike} 
L. Deike, D. Fuster, M. Berhanu, and E. Falcon, Phys. Rev. Lett. {\bf 112}, 234501 (2014).
\bibitem{denis}
P. Denissenko,  S. Lukaschuk, and S. Nazarenko, Phys. Rev. Lett. {\bf 99}, 014501 (2007). 
\bibitem{denis1}
S. Lukaschuk, S. Nazarenko, S. McLelland, and P. Denissenko, Phys. Rev. Lett. {\bf 103}, 044501 (2009)
\bibitem{Falcon}
E. Falcon, C. Laroche, and S. Fauve, Phys. Rev. Lett. {\bf 98}, 094503 (2007). 
\bibitem{aubourg} 
Q. Aubourg, and N. Mordant, Phys. Rev. Lett. {114}, 144501 (2015). 
\bibitem{lvov03} 
Y. Lvov, S.N. Nazarenko, and R. West, Physica D {\bf 184}, 333 (2003). 
\bibitem{Dyachenko}
S. Dyachenko, A.C. Newell, A.N. Pushkarev, and V. E. Zakharov, Physica D {\bf 57}, 96 (1992). 
\bibitem{Galtier2003}
S. Galtier, Phys. Rev. E {\bf 68}, 015301 (2003);  P. Mininni, and A. Pouquet, Phys. Fluids {\bf 22}, 035106 (2010).
\bibitem{Galtier2014} 
S. Galtier, J. Fluid Mech. {\bf 757}, 114 (2014). 
\bibitem{chibbaro} 
S. Chibbaro, and C. Josserand, Phys. Rev. E {\bf 94}, 011101 (2016).
\bibitem{space} 
S. Galtier, S.V. Nazarenko, A.C. Newell, and A. Pouquet, J. Plasma Phys. {\bf 63}, 447 (2000);  
S. Galtier, J. Plasma Phys. {\bf 72}, 721 (2006); R. Meyrand, S. Galtier, and K.H. Kiyani, Phys. Rev. Lett. {\bf 116}, 105002 (2016).
\bibitem{maggiore}
M. Maggiore, {\em Gravitational Waves I: Theory and Experiments}, (Oxford University Press, Oxford, 2008). 
\bibitem{weinberg}
S. Weinberg, {\em Gravitation and Cosmology: Principle and Applications of the General Theory of Relativity}, (Wiley, New York, 1972). 
\bibitem{HZ14}
Y. Hadad, and V. Zakharov, J. Geom. Phys. {\bf 80}, 37 (2014).
\bibitem{schw}
K. Schwarzschild, K.P. Akad. Wiss {\bf 1}, 189 (1916).
\bibitem{witten}
E. Witten, Nucl. Phys. B {\bf 311} 46 (1988). 
\bibitem{guth}
A.H. Guth, Phys. Rev. D {\bf 23}, 347 (1981). 
\bibitem{colm}
C. Connaughton, and S.N. Nazarenko, Phys. Rev. Lett. {\bf 92}, 044501 (2004).
\bibitem{lacaze}
R. Lacaze, P. Lallemand, Y. Pomeau, and S. Rica, Phys. D.  {\bf 152-153}, 779 (2001). 
\bibitem{caprini}
C. Caprini, J. Phys. Conf. Ser. {\bf 610}, 012004 (2015).
\bibitem{janssen}
G. Janssen et al., AASKA14 {\bf 37}, 1 (2015).
\end{thebibliography}
\end{document}